# Ranking Online Consumer Reviews

Sunil Saumya[1]

*National Institute of Technology Patna, Bihar, India*

Jyoti Prakash Singh[2]

*National Institute of Technology Patna, Bihar, India*

Abdullah Mohammed Baabdullah[3]

*King Abdulaziz University, Jeddah, Kingdom of Saudi Arabia*

Nripendra P. Rana[4]

*Swansea University Bay Campus, Swansea, UK*

Yogesh k. Dwivedi[5]

*Swansea University Bay Campus, Swansea, UK*

**Abstract**

The product reviews are posted online in the hundreds and even in the thousands for some popular products. Handling such a large volume of continuously generated online content is a challenging task for buyers, sellers and even researchers. The purpose of this study is to rank the overwhelming number of reviews using their predicted helpfulness score. The helpfulness score is predicted using features extracted from *review text data*, *product description data* and *customer question-answer data* of a product using random-forest classifier and gradient boosting regressor. The system is made to classify the reviews into low or high quality by random-forest classifier. The helpfulness score of the high quality reviews is only predicted using gradient boosting regressor. The helpfulness score of the low quality reviews is not calculated because they are never going to be in the top *k* reviews. They are just added at the end of the review list to the review-listing website. The proposed system provides fair review placement on review listing pages and making all high quality reviews visible to customers on the top. The experimental results on data from two popular

[1]   Sunils.cse15@nitp.ac.in
[2]   jps@nitp.ac.in
[3]   baabdullah@kau.edu.sa
[4]   nrananp@gmail.com
[5]   ykdwivedi@gmail.com

Indian e-commerce websites validate our claim, as 3-4 new high-quality reviews are placed in the top ten reviews along with 5-6 old reviews based on review helpfulness. Our findings indicate that inclusion of features from *product description data* and *customer question-answer data* improves the prediction accuracy of the helpfulness score.

**Keywords**: E-commerce; Big Data Challenge; Machine Learning; Helpfulness; Online Generated Reviews

**1. Introduction**

The online product review is a way for business managers and customers to connect with other customers (Heinonen, 2011). As with other forms of social media, online review sites provide business managers and customers the opportunity to get connected on a platform (Heinonen, 2011). According to brightlocal.com, 80% of customers read reviews about a product before making a purchase decision (BrightLocal, 2016). For a single product, reviews are posted in the hundreds and sometimes even in the thousands. Thus, it is difficult for any user to go through all the reviews before making a purchase decision. The same survey of brightlocal.com also states that 90% of consumers read ten or less than ten reviews before making their purchase decisions (BrightLocal, 2016). Hence, the listing of reviews plays an important role. Putting high-quality reviews on the top can minimize users' time because they can get their information from very few quality reviews. Generally, reviews for almost all e-commerce websites are sorted based on recency or helpfulness. Helpful reviews are those that have obtained positive votes from similar users[6]. Similarly, the most recent reviews are presented as a list of reviews that are sorted based on the time at which the review was posted (Ghose and Ipeirotis, 2011). Amazon has recently introduced a new sorting mechanism referred to as its top reviews. At the top, further filtration of the reviews can be done based on categories such as all positive reviews[7], all negative reviews[8], or reviews with star ratings. However, the sorting algorithm for top reviews used by Amazon is not public.

---

[6] Similar users refer to users who have brought the same product or services.

People prefer to read helpful reviews over recent reviews because helpful reviews indicate sufficient acceptance by similar users, which helps establish trust in the review. The helpfulness of online reviews is achieved by asking a simple question: "was this review helpful to you?" Users are asked to give their opinion by clicking either "thumbs up" or "thumbs down" buttons. However, due to a large number of reviews available for each product, helpfulness voting does not solve all problems. In fact, the listing of reviews based on helpfulness votes suffers from the Matthew effect (Merton, 1968), which states that a review that is at the top of the review list will remain on top because users mostly access and vote for a few of the top reviews before making their purchase decisions, whereas a review that is at the bottom of the review list will remain at the bottom because users hardly notice them (Singh et al., 2017a). A new review that has just appeared on the product page and has not received significant votes until recently may remain at the bottom of the review list. It may be the case that the review has the potential to rise to the top of the review list. This may negatively affect the relationship between reviewer and website. The reviewer may become demotivated and may not write reviews for other products in the future. Hence, there is a need for sorting reviews based on content to place them in the proper place to mitigate the Matthew effect and to obtain fair visibility.

Singh et al. (2017a) raised this issue and proposed a model to automatically predict the helpfulness of online consumer reviews using only review text. Similarly, (Ghose and Ipeirotis, 2007) proposed a novel review ranking system using subjectivity analysis techniques. However, they did not mention how to place a review in its proper place and evaluate it.

---

[7] A review with star rating 4 or 5 is termed as positive review.
[8] A review with star rating 1 or 2 or 3 is termed as negative reviews.

In this article, we extend the work of (Singh et al. 2017a) by constructing a better regression prediction model to place a review in its proper place in the review list based on predicted helpfulness score. To the best of our knowledge, most of the works undertaken to date regarding helpfulness prediction have mainly focused only on finding the best-features set from the review text and its meta-data. Thus, along with finding a good set of features, the current research proposes the idea of building a review helpfulness prediction regression model with a classifier. That is, we propose using a classifier to identify reviews as either low- or high-quality, and we use only the high-quality reviews to build the regression model. High-quality reviews are those that have obtained sufficient acceptance by similar users in the form of helpfulness votes, whereas low-quality reviews are those that do not have sufficient information to convince the buyers. Usually, low-quality reviews present the information in a few words or a few sentences, which has little impact on buyers. Hence, such reviews do not receive any helpful votes. Generally, when we build a regression model, we include data with both low and high-quality reviews (Chua and Banerjee, (2015); Huang et al., (2015); Ngo-Ye and Sinha, (2014)). However, training the proposed regression model on only high-quality reviews will result in better performance because low-quality reviews cannot fit into the idea of predicting high-quality reviews. Hence, we intend to remove such low-quality reviews while training the regressor. The removal of low-quality reviews also improves the time complexity because there will be no list change for low-quality reviews.

A high-quality (or helpful) review is likely to possess other aspects of products, such as *product description data* and *customer question-answer data*, in a review sentence. *Product description data* are the structural aspects of the product information listed by an e-commerce retailer on the product listing webpage, for example, a memory card's *capacity* or a camera's *megapixels*. People generally match their requirements with the description of a product that is available in the form of *product description data*, and then they read the reviews to validate such descriptions. A helpful review dispenses as many measurable contents as possible to provide detailed information about the *product description data* (Kim et al., 2006; Zheng et al., 2013; Karmaker et al., 2016). For example, a high-quality mobile phone review tends to comprehensively elaborate product aspects, including battery power, operating system version, camera features, and internal memory capacity. It also provides exhaustive personal opinions and experiences, which are not reflected in low-quality reviews. In a manual inspection of more than 100 products, it was found that positive helpful reviews, which include product descriptions, received more votes than did other reviews that do not include such information. However, the critical reviews (negative) are found to be an exception to this. Hence, a fundamental requirement is to investigate reviews at the feature level. This requires i) the recognition of product features and ii) the association of each review with one or more recognized feature. The recognition of product features can be done directly by extracting *product description data* from the product web page (see Figure 1). The second job of associating each review with product description data is relatively challenging. In this paper, we use a similarity measure to identify the key features mentioned in a review sentence and to measure how this affects high-quality review prediction.

**Technical Details**

| | |
|---|---|
| OS | Android |
| RAM | 3 GB |
| Item Weight | 163 g |
| Product Dimensions | 15.1 x 0.8 x 7.6 cm |
| Batteries: | 1 Lithium ion batteries required. (included) |
| Item model number | 6X |
| Wireless communication technologies | Bluetooth, WiFi Hotspot, WiFi Bridge, Wi-Fi Direct, Wi-Fi WPS |
| Connectivity technologies | GSM, (850/900/1800/1900 MHz), 3G, WCDMA, (1/5/6/8/19), 4G, FDD, (1/3/5/7/8/19/28), GPRS, EDGE, WiFi, VoLte |
| Special features | Dual SIM, GPS, Music Player, Video Player, FM Radio, Hall effect sensor, Fingerprint sensor, Proximity sensor, Ambient light sensor, eCompass, Gravity sensor, Phone status indicator, E-mail |
| Other camera features | 8MP |
| Form factor | Touchscreen Phone |
| Weight | 163 Grams |
| Colour | Grey |
| Battery Power Rating | 3340 |
| Phone Talk Time | 23 Hours |
| Phone Standby Time (with data) | 650 Hours |
| Whats in the box | Handset, Charger, USB Cable, Quick Start Guide and Ejection Tool |

Figure 1. A sample example of *product description data* of a product from Amazon

To obtain more customer engagement, e-commerce websites initiated customer *question-answer sections* related to a particular product along with reviews of that product. *Customer question-answer* data usually contain a collection of questions that are frequently discussed by customers in their review text. Websites such as Amazon.in and Snapdeal.com provide the ability to ask any question about product features and services. Other verified users[9] can answer such questions based on their experience. Users can also discuss these questions in their review texts. It has been observed that a review obtains more acceptance by other users if it includes the topics on which questions are asked. Hence, the similarity between the customer question-answer data and review text may play a significant role in predicting the helpfulness of the review.

Based on the above observations, we have formulated three research questions as follow:

*RQ1: Will excluding low-quality reviews as data lead to the improved performance of the regression model?*

*RQ2: Does the inclusion of product description data in online consumer reviews affect the helpfulness of such reviews? If so how?*

*RQ3: Does the inclusion of topics on question-answer data affect the helpfulness of online consumer reviews? If so how?*

In the present research, we have evaluated the important features and their effects on the prediction of review helpfulness. A machine-learning-based automated system is developed to build a better regression prediction model to organize review listing. The data imbalance problem is also addressed before applying the machine learning algorithms to improve the results. To our knowledge, none of the existing studies have considered the issue of the data imbalance problem, which is inherent between high-quality and low-quality reviews. Predicting the helpfulness votes of reviews based on an imbalanced dataset will affect a system's accuracy. We address the imbalanced dataset issue after labelling to achieve a better classifier.

---

[9] The reviewer who has purchased the product is known as a verified user.

The rest of the article is organized as follows: We briefly discuss the recent relevant work done in the area of review helpfulness in Section 2. Section 3 describes our methodology, including the data collection, feature extraction and machine learning algorithms used in the present research. The results are presented in Section 4, followed by the discussion in Section 5. Finally, we conclude the paper with some future directions in Section 6.

**2. Literature Review**

Due to the immense success of e-commerce, a great deal of research has been conducted to enhance customer engagement. One of the primary ways to engage customers is to provide them with the opportunity to describe their experience in the form of reviews on the product listing page. The question of how to determine the helpfulness of such reviews has attracted a number of researchers over the years. Researchers have used different approaches to predict the helpfulness of reviews. For example, some researchers have used a machine learning approach for this task, while others have used a statistical approach toward the same ends. In this section, we review some of the important contributions of such research to align our proposed work with it and identify a gap in the literature that will validate our motivation to conduct the current study. We present the previous works in two categories: i) an automated approach for helpfulness prediction and ii) manual approach for helpfulness prediction.

*2.1 Automated approach for helpfulness prediction:*

In an automated approach, most of the earlier works used either a supervised machine learning technique or a statistical technique for review helpfulness prediction. For example, Chua and Banerjee (2016) and Lee and Shin (2014) used a statistical approach to establish the relationship between review helpfulness and review sentiment and between product type and information quality. Chua and Banerjee (2016) used amazon.com reviews to validate their model. They found that review helpfulness was independent of product type. They also found that review helpfulness varies as a function of review sentiment. Mudambi and Schuff (2010) explored the factors that make customer reviews helpful. Based on a paradigm of search and experience goods, they found that review extremity and review depth significantly affect the helpfulness of online reviews. By analyzing 1,587 reviews of six product types such as mp3 players and digital cameras, they concluded that, for experience goods, moderate ratings more significantly affect helpfulness than extreme ratings. However, for search goods, review depth showed a better association. They found that review extremity, review depth, and product type affected the perceived helpfulness of a review. Racherla and Friske (2012) examined the relationship between the determinants of review helpfulness and different product types using optimal least square regression. The types of product discussed were search, experience and credence. Analyzing 3000 reviews from Yelp.com, they found that, for experience products, reviewer expertise and reputation affected review helpfulness positively. However, for both search and experience products, review valance showed a convex-shaped relationship with usefulness.

Lee and Choeh (2014) proposed a helpfulness prediction model using a neural network regression model. They used an amazon.com dataset for validation. They used determinants such as product type, review characteristics and textual characteristics. They extracted 20 features and found that the product type and textual characteristics are important determinants for review helpfulness when predicting reader's perception. Liu et al. (2007) addressed the problem of detecting low-quality reviews using a classification-based approach (Singh et al., 2017b). Three aspects of product reviews were explored, specifically, informativeness, subjectiveness, and readability. A set of rules for judging the quality of reviews was defined. The proposed approach first distinguished a low-quality review from a high-quality review and, hence, the opinion summarization result was obtained based only on the high-quality reviews. Qazi et al. (2016) focused on a concept-level approach to analyze the helpfulness of online reviews. They considered both the qualitative and quantitative aspects while building their model for helpfulness prediction. They defined three types of review: regular reviews, comparative reviews and suggestive reviews: a regular review is often referred to simply as an opinion; a comparative review expresses a relation of similarities or differences among two or more products or services; and a suggestive review is defined as directing someone to do something in a polite manner. In that study, Qazi et al. (2016) used a tobit regression model based on 1500 reviews of different hotels from TripAdviser for their analysis. Their results suggested that the review type, the number of concepts in a review and the average number of concepts per sentence have a positive significant effect on review helpfulness. The regular reviews did not have a significant effect on helpfulness. They also found that review types such as regular, comparative and suggestive reviews significantly affected purchase decision and business policy.

Chua and Banerjee (2015) and Chen and Huang (2013) investigated the interplay among reviewer reputation, review depth and review rating. By analyzing Amazon data, they found that reviewer profile and review depth were positively associated with review helpfulness, whereas review rating was negatively associated with review helpfulness. Chua and Banerjee (2015) used multiple regression, specifically a tobit estimation, to validate their model. Ngo-Ye and Sinha (2014) developed and compared several text regression models for predicting review helpfulness using reviewer engagement characteristics such as reputation, commitment and current activity. They used datasets of Yelp and Amazon to conduct their experiments. Liu et al. (2008) presented a non-linear regression model for helpfulness prediction using three important factors: the reviewer's expertise, the writing style of the review and the timeliness of the review. They used IMDB movie reviews to test the proposed model. Krishnamoorthy (2015) proposed textual linguistic features along with a review metadata-features-based approach for predicting the helpfulness of online consumer reviews. They used two real datasets from (Blitzer, Dredze, and Pereira, 2007) and Amazon for the experimentation. Some of the linguistic features were marked as effective for predicting review helpfulness. Weathers et al. (2015) focused on the diagnosticity and credibility of electronic word-of-mouth. Diagnosticity explains the uncertainty and equivocality of the reviewer, whereas credibility defines the trust and expertise of the reviewer. They studied amazon.com product reviews and found that review writers who explicitly attempt to enhance review diagnosticity or credibility are often ineffective or systematically unhelpful.

Ghose and Ipeirotis (2011) examined the impact of online reviews on product sales and on perceived helpfulness. They used a random forest-based classifier to predict the usefulness of the reviews and their impact on sales. They assessed the relative impact of reviewer-related features, review subjectivity features and review readability features. They observed that reviews of medium length with few spelling errors were more helpful compared to very short and very long reviews with spelling errors. They also found the linguistic feature to be an important feature for product sales. Huang et al. (2015) suggested that the effect of word count on review helpfulness has some threshold. Beyond that threshold, the effect diminishes significantly. They used tobit regression on amazon.com reviews for their analysis. They found that review framing was an important predictor for review helpfulness. They also found that reviewer experiences and their impact were not statistically significant predictors of helpfulness but that past helpfulness records tended to predict future helpfulness ratings. Kim et al. (2006) assessed the helpfulness of online reviews automatically using support vector machine (SVM) regression. An amazon.com dataset was used to evaluate the proposed idea. They found that the length of a review, unigrams, and product rating were important features for assessing helpfulness. Siering and Muntermann (2013) extended the understanding of review diagnosticity based on review sentiment, product quality, and review uncertainty (Saumya et al., 2016). Review diagnosticity basically addresses review extremity and review depth. They used tobit regression to predict review helpfulness.

Zhu et al. (2014) developed an integrative approach to estimate review helpfulness using reviewer credibility, rating extremity and service price. They used econometrics regression analysis on hotel reviews to evaluate the proposed model. They found that reviewer expertise and reviewer popularity had a positive effect on perceived helpfulness. They also found that product price differently affects perceived helpfulness. Malik and Hussain (2017) built a helpfulness predictive model and investigated the effect of emoticons' contribution to helpfulness. They used a deep learning neural network model to implement the presented model. The results indicated that when an individual feature category is considered, positive emotion features are the best predictors. However, negative emotion features and visibility features obtain comparable performance. Korfiatis et al. (2012) explored the interplay between review helpfulness and rating (Tsang and Prendergast, 2009) and review helpfulness and readability, and among all three together. They used tobit regression on amazon.com for analysis. They constructed the model based on three elements: conformity, understandability and expressiveness. The main findings of this research were as follows: (i) the helpfulness of a review is directly affected by its star rating, its qualitative characteristics and, in particular, its review readability and (ii) the review length is affected by the review rating score.

*2.2 Manual approach for helpfulness prediction:*

In the manual approach, researchers have used a laboratory approach or a survey-based approach for predicting the helpfulness of reviews. Mackiewicz and Yeats (2014) examined the potential characteristics of reviews including credibility, informativeness, and readability to obtain a better mode of assessing review quality. They conducted an audience survey to gauge the quality of products on a five-point scale. They found informativeness to be the most important component of review quality. Allahbakhsh et al. (2015) measured the reviewer trust ranking and the rating score. They used the term trust to represent the community judgment on the quality of a person, whereas the term rating score refers to the overall quality of a product from the community point of view. Li et al. (2013) conducted a laboratory experiment to evaluate the helpfulness of online product reviews. They also evaluated the impact of source- and content-based review features on review helpfulness. The findings showed that the consumer perceives customer-written product reviews to be more helpful than those written by an expert. They also perceive concrete reviews to be more helpful than abstract reviews. A review with a low level of content abstractness yields the highest perceived review helpfulness. Baek et al. (2015) explored the association between review extremity and review helpfulness from a normative social influence perspective. They used reviews from amazon.com and yelp.com for analysis. Their findings suggest that users tend to consider reviews helpful when their rating is close to the average rating of products.

In all of the existing studies that have addressed high-quality reviews, researchers have either used classification to separate reviews into high-quality or low-quality, or they used regression to predict a set of the best of all of the subject reviews. In line with these studies, the current research tries to predict the set of top (high-quality) reviews by integrating both classification and regression into one model. We first classify reviews into high- and low-quality using classifiers and then feeding only high-quality reviews to the regressor to predict the set of best reviews. All of the earlier works used aspects such as product type, review extremity, review text, reviewer reputation, and/or sales as determinants of a helpfulness prediction. However, none of them have considered *customer question-answer* data to evaluate review helpfulness. This research shows that features from *product description data,* and *customer question-answer data* have a strong effect while ranking reviews based on their helpfulness prediction.

## 3. Research Methodology

The major workflow of the current research is explained in Figure 2. The major components of the system are (i) Data collection, (ii) Data cleaning, (iii) Feature extraction, (iv) Classification and (v) Rank prediction.

*3.1. Data collection*

We collected data from two popular Indian e-commerce websites, namely, (i) Amazon.in, and (ii) Snapdeal.com. As there is no Application Programming Interface (API) available for these websites, data crawling technique is used to crawl webpages from these websites. The required information is extracted from those crawled pages. The collected data are grouped into three categories: (i) *product description data*, (ii) *customer question-answer data*, and (iii) *review text* for each product. For *product description data*, we collected the textual details of the product that were available on the product listing webpage and kept them in a table called *product description data*. All the available *customer question-answer data* for the product are collected and placed in another table called *question-answer data*. Finally, all the reviews of the said product are extracted and placed in another table called *review data*. The *review data* contains the following fields: product ID, review date, review heading, rating, review text and helpful votes (Saini et al., 2017).

A total of 29,215 and 12,686 review data are collected from amazon.in (Amazon) and snapdeal.com (Snapdeal), respectively. The data collection is done for five different products: (i) power bank, (ii) mobile phone, (iii) memory card, (iv) book and (v) baby product. The data were collected during the months from October 2016 to December 2016. For the power bank, we collected 3,371 reviews from Amazon, of which 1,971 reviews were positive and the remaining 1,400 reviews were negative. The total number of questions discussed for the power bank on Amazon was 861. We collected 3,481 positive and 1,550 negative reviews for the mobile phone from Amazon. The number of questions discussed for the mobile phone was 784. Similarly, from Amazon, we collected 12,071 positive and 2,000 negative reviews, 1,101 positive and 200 negative reviews, and 4,241 positive and 1,200 negative reviews for its products, specifically the memory card, the baby product and the book, respectively. We also collected 1,000 and 28 questions discussed for the memory card and the baby product from Amazon.in. We did not find any questions about the book on Amazon. The number of negative reviews of all five products on Snapdeal was much less in comparison with its positive reviews. For the power bank, the number of positive and negative reviews was 1,365 and 65, whereas the total number of questions asked was 648. The number of positive and negative reviews collected for the mobile phone and the memory card from Snapdeal were 2,005 and, 26 and 8,869 and 82, respectively. The number of negative reviews and *questions-answers* was not present for the baby product and the book. The number of positive reviews collected for these two products were 111 and 163, respectively. A detailed description of the data collected is provided in Table 1.

Table 1. Number of reviews and question-answers collected for each product

| Product | Amazon | | | Snapdeal | | |
|---|---|---|---|---|---|---|
| | Number of Positive reviews | Number of negative reviews | Number of question-answer | Number of Positive reviews | Number of negative reviews | Number of question-answer |
| Power bank | 1,971 | 1,400 | 861 | 1,365 | 65 | 648 |
| Mobile phone | 3,481 | 1,550 | 784 | 2,005 | 26 | 320 |
| Memory card | 12,071 | 2,000 | 1,000 | 8,869 | 82 | 477 |
| Baby product | 1,101 | 200 | 28 | 111 | 0 | 0 |
| Book | 4,241 | 1,200 | 0 | 163 | 0 | 0 |

*3.2. Data cleaning*

The review and *product description* and *question-answer data* contain some Unicode characters or images of the products. We remove all Unicode characters and images from our data using a Python program. Some of the reviews obtained more than 1000 votes, and most of the reviews received votes in the range of 0-10. We curtail the votes of these exceptionally high vote reviews to triple of the non-zero average helpful reviews to ensure that the learning is smooth. After the data cleaning, we were left with 27,190 reviews from Amazon and 10,255 reviews from Snapdeal.

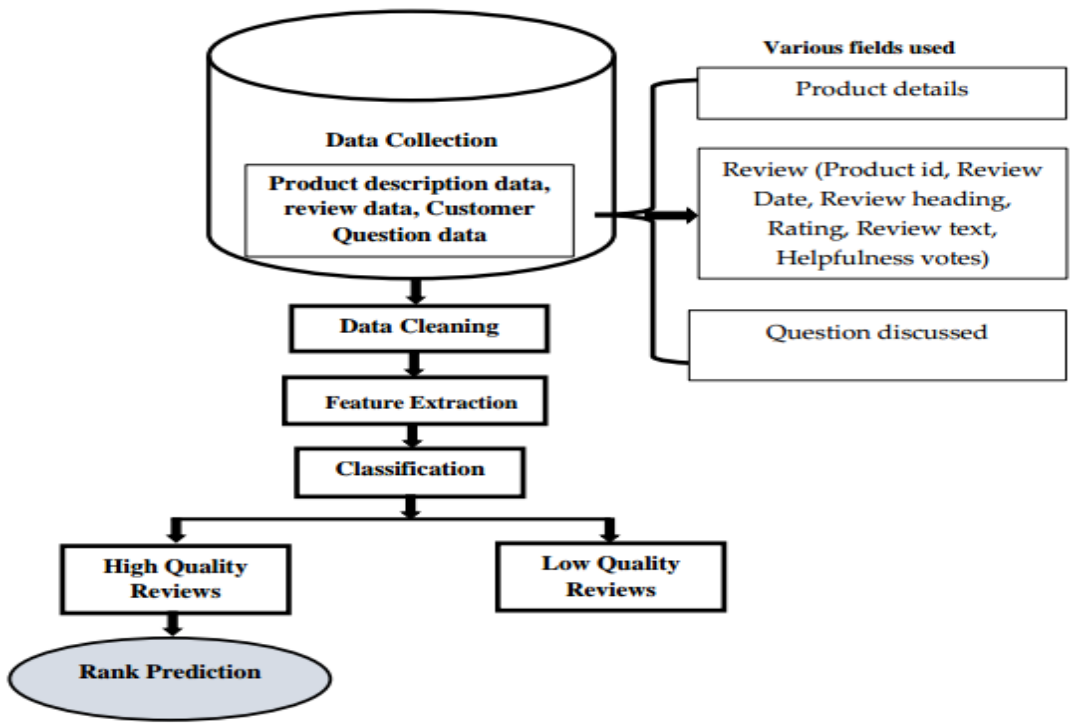

Figure 2. Flow diagram of proposed system

3.3. Feature extraction

We extracted 17 features from three types of data used for this research. A total of 15 features are extracted from the review data (Ghose and Ipeirotis, 2011; Lee and Choeh, 2014; Roy et al., 2018; Singh et al., 2017a). These are *(i) Noun, (ii) Adjective, (iii) Verb, (iv) Flesch_reading_ease, (v) Dale_chall_reading, (vi) Difficult_words, (vii) Length, (viii) Set_length, (ix) Wrong_words, (x) One_letter_words, (xi) Two_letter_words, (xii) Longer_letter_words, (xiii) Lex_diversity, (xiv) Entropy, and (xv) Rating*. A detailed description of each feature is provided in Table 2. In addition, two more features are extracted from the *product description data* and *customer question-answer data*. From the *product description data*, we extracted one feature, *(xvi) Desc_sim. Desc_sim* calculates the similarity between the *product description* and the review text. From the *customer question-answer data*, we extracted one feature, *(xvii) QA_sim*. *QA_sim* evaluates the similarity between the question discussed on the product page and the review text. Helpfulness vote is used as the dependent variable.

We used cosine similarity to calculate the *Desc_sim* and *QA_sim*. Generally, cosine similarity is used as a similarity measure in document comparisons. We considered *product description* as document $d_1$ and the review text as document $d_2$. For each document, we derive a vector. We use a '*Bag of words*' representation of the vector where each term is represented using a *one hot encoder*. The set of documents in a collection is then viewed as a set of vectors in a vector space. Each term in a document will have its own axis. Using the formula given below, we can determine the similarity between any two documents.

$$\text{Cosine Similarity } (d_1, d_2) = \frac{d1.d2}{||d1||*\sqrt{d2v}} \quad (1)$$

Cosine similarity measures the two documents on the scale of 0 to 1, where 0 refers to "no match," and 1 refers to "100% match". For example, we took a sample review of a product power bank and its description from Amazon.

Description ($d_1$):

*"Brand - Intex, Model ID - IT-PB11K, Compatible For - Micro USB, Capacity - 11000 mAh, Color - White, Power Input - DC5V / 2.1A, Power Source - Mini USB, Output Power - 5V 1A, 5V2A \and 5V 2A, LED Indicator – Yes".*

Review ($d_2$):

*"I love this product. 11000mah power bank is best. i have been using this product for last three months and its amazing. Input Power: 10.5W [5V / 2.1A] - USB Micro-B. Output Power: USB Type A sockets power output below. Socket 1 - 5W [5V / 1A], Socket 2 - 10.5W [5V / 2.1A], Socket 3 - 10.5W [5V / 2.1A], if i will charge my mobile with original charger it takes 3 hours and if i will charge my mobile with power bank it takes approx.... 2.2 hours."*

Using Equation 1, we calculated the *Desc_sim* between *product description* ($d_1$) and review ($d_2$). The similarity was found to be 0.1869231095. Similarly, we took sample questions with regard to a product power bank from its *customer question-answer* section as shown in Figure 3. We kept the questions in one document $d_3$.

Customer questions ($d_3$):

*"how much time he need chage the 2500mh batter phone? Are three USB cables provided with this? Is it compatible for i phone?? also tell me, can i charge my i phone when power bank is charging?? Is it compatible for microsoft lumia 535??"*

![Figure 3 sample Q&A screenshot]

Figure 3. A sample question-answer of power bank from Amazon

Using Equation 1, we calculated the *QA_sim* between document $d_2$ and $d_3$. The similarity was calculated as 0.2518. Similarly, we calculated *Desc_sim* and *QA_sim* for each review of Amazon and Snapdeal.

*3.4. Classification*

Most of the reviews in the online product review list do not receive any votes because of their incompleteness and lack of informativeness. We sought to remove such reviews because they cannot fit our proposed idea of predicting the top *k* reviews based on their helpfulness votes. To achieve this, we grouped our dataset into two classes: (i) Class 1: *high-quality review* and (ii) Class 0: *low-quality review*. Usually, low-quality reviews contain the information in a few words or a few sentences that create any little impact on buyers. Hence, these reviews do not receive helpfulness votes.

Table 2. List of independent variables and their description

| Variable type | Feature | Description |
|---|---|---|
| Review data | Noun | The number of nouns in the review text |
| | Adjective | The number of adjectives in the review text |
| | Verb | The number of verbs in the review text |
| | The Flesch_reading_ease | The Flesch Reading Ease Score. The following scale is helpful in assessing the ease of readability in a document: 90-100: Very Easy, 80-89: Easy, 70-79: Fairly Easy, 60-69: Standard, 50-59: Fairly Difficult, 30-49: Difficult, 0-29: Very Confusing. |
| | Dale_chall_RE | The DaleChall readability formula is a readability test that provides a numeric gauge of the comprehension difficulty that readers encounter when reading a text. It uses a list of 3000 words that groups of fourth-grade American students could reliably understand, considering any word not on that list to be difficult. |
| | Difficult_words | Difficult words do not belong to the list of 3000 familiar words. |
| | Length | Total words in the review. |
| | Set_length | Total unique words in the review. |
| | Wrong_words | Words that are not found in Enchant English dictionary. |
| | One_letter_words | The number of one-letter words in the review. |
| | Two_letter_words | The number of two-letter words in the review. |
| | Longer_letter_words | The number of more than two-letter words in the review. |
| | Lex_diversity | The ratio of unique words to total words in the review. |

|  | Entropy | The entropy indicates how much information is produced on average for each word in the text. |
|---|---|---|
|  | Rating | The star rating of the product. |
| Product description data (Proposed variable) | Desc_sim | Similarity between product description and review text. |
| Customer question-answer data (Proposed variable) | QA_sim | Similarity between questions discussed for a product and its review text. |

To label the reviews into two classes, we converted the continuous variable helpfulness into a binary one. For this, we can simply select a *threshold* value *t* and mark all reviews > *t* as high-quality reviews and the rest as low-quality reviews. However, selecting the *threshold* value is slightly trickier. As the number of helpfulness votes received by any review varies by product, selecting the *threshold* value *t* randomly would not be the right approach. For example, setting the *threshold t* too high would mean that some of the high-quality reviews would be labelled low-quality, and setting the value *t* too low would mean that some low-quality reviews would fall in the high-quality category. To select an optimal value of *t*, we first calculated the average helpfulness of each product. Our analysis indicated that by selecting the average helpfulness value as a *threshold* value, we can classify the data set into high-quality reviews and low-quality reviews more accurately. Hence, if a review received helpfulness votes > *threshold t*, we label it as a high-quality review. Otherwise, the review is labeled a low-quality review.

After labeling the reviews into two classes, we use classification algorithms to train our models with the labelled data. The learned model can be further used to classify the unseen reviews as high-quality reviews or low-quality reviews. Three different classification approaches, support vector machine (SVM), naïve Bayes (NB) and random forest (RF), are used for the experiments. We designed four conceptual models for classification: (i) Case 1 (or baseline model): classification without including the *product description data* and the *customer question data*, (ii) Case 2: classification with the *product description data*, (iii) Case 3: classification with the *customer question-answer data*, and (iv) Case 4: classification with both the *product description* and the *customer question-answer data*. We refer to Case 1 as a baseline model as we compare our other three cases with it to check the impact of the *product description data* and the *customer question answer data* on a classifier's accuracy. In all four cases, we include all other 15 features extracted from the review data.

*3.5. Rank prediction*

We use two different regression techniques, linear regression and gradient boosting (GB) regression to fit a line for rank prediction. We divide our entire dataset into two parts, a training set and a testing set. The ratio of the training set to the testing set is 3:1. We conduct the rank prediction in two different parts: (i) rank prediction with classifier, and (ii) rank prediction without classifier. In the rank prediction with classifier, we first classify the reviews into high-quality reviews and low-quality reviews as discussed in Section 3.4. Then, we fed high-quality reviews to the regressor, which gives the top *k* reviews among high-quality reviews based on its predicted helpful vote. However, in the rank prediction without classifier, we directly fed all our data (without any classification) to the regressor, which gives the top *k* reviews among the total reviews based on the predicted helpful vote.

**4. Result**

As discussed in Section 3.5, we conducted the whole experiment in two parts: (i) rank prediction with classifier and (ii) rank prediction without classifier. The detailed results of both parts are discussed in the following subsections. The factors of precision, recall, $F_1$-score and receiver operating characteristics (ROC) curve (Davis and Goadrich, 2006) were used to measure the performance of the classifier, whereas mean square error (MSE) was used to measure regressor performance.

A precision is the ratio of a number of events one can correctly recall to the number of all events one recalls (mix of correct and wrong recollections). In our case, it is the ratio of the number of high-quality reviews we correctly classified to all the classified reviews (a mix of high-quality and low-quality reviews). Mathematically, precision can be represented as

*A number of high-quality reviews we correctly classified = True positive (they are high-quality reviews and we classified them as high quality)*

*A number of all classified reviews = True positive (they are high-quality reviews and we classified them as high quality) + False positive (they are low-quality reviews but we classified them as high-quality reviews)*

$$Precision = \frac{TruePositive}{TruePositive + FalsePositive} \quad (2)$$

Recall is the ratio of a number of events one can correctly recall to a number of all correct events. In our case, it is the ratio of correctly classified high-quality reviews to all high-quality reviews.

*A number of high-quality reviews we correctly classified = True positive (they are high-quality reviews and we classified them as high quality)*

*A number of all high-quality reviews = True positive (they are high-quality reviews and we classified them as high quality) + False negative (they are high-quality reviews but we classified them as low quality)*

$$Recall = \frac{TruePositive}{TruePositive + FalseNegative} \quad (3)$$

The $F_1$-score is the harmonic mean of Precision and Recall. Mathematically, the F-score can be calculated as

$$F_1\text{-score} = \frac{2*Precision*Recall}{Precision+Recall} \quad (4)$$

An ROC (receiver operating Characteristic) curve is a graphical plot that measures the performance of a binary classifier. The curve is created by plotting the true positive rate (TPR) against the false positive rate (FPR) at various threshold settings. The classifier performance is more accurate if the curve follows the left-hand border and then the top border of the ROC space (Davis and Goadrich, 2006).

In regression, mean square error (MSE) is a measure of how close a fitted line is to the data points. It is the squared difference between the values predicted by a model and the values actually observed. The root mean square error is simply a square root of the MSE.

*4.1. Rank prediction with classifier*

We checked the dimensions of the reviews after labeling them into high- and low-quality reviews. We found that the data were highly imbalanced between the two classes. Of the 27,190 Amazon reviews, only 2,232 reviews were labeled Class 1 (high-quality review), and the remaining 24,958 reviews were labeled Class 0 (low-quality reviews). To address this, we used the *SMOTE* (Chawla et al., 2002) over-sampling technique to balance the data over the two classes. Finally, the balanced set of reviews was given to the classifiers as an input. We divided the dataset for training and testing in a ratio of 3:1. In our experiment, we used three classifiers, SVM, NB, and GB, to validate our baseline model (or Case 1 model). The detailed results of Case 1 for testing the model are shown in Table 3. As seen from Table 3, the RF classifier performed better, with an $F_1$ score of 0.86 for our target class 1, than did the other two approaches. Hence, we tested the other three comparative conceptual models, *Case 2, Case 3,* and *Case 4*, only with the RF classifier. The testing results of all three cases are shown in Table 4.

Table 3. Classification accuracy result of *Baseline conceptual model* for test data of Amazon

| Approaches | Class 0 | | | Class 1 | | |
|---|---|---|---|---|---|---|
| | Precision | Recall | $F_1$-score | Precision | Recall | $F_1$-score |
| Naive Bayes | 0.57 | 0.93 | 0.70 | 0.81 | 0.29 | 0.43 |
| Support Vector Machine | 0.82 | 0.78 | 0.80 | 0.79 | 0.83 | 0.81 |
| Random Forest | 0.86 | 0.91 | 0.88 | 0.87 | 0.86 | **0.86** |

Table 4. RF Classification accuracy result of *Case 2, Case 3, and Case 4 conceptual models* for test data of Amazon

| | Class 0 | | | Class 1 | | |
|---|---|---|---|---|---|---|
| | Precision | Recall | $F_1$-score | Precision | Recall | $F_1$-score |
| Case 2 | 0.94 | 0.86 | 0.90 | 0.88 | 0.95 | 0.91 |
| Case 3 | 0.89 | 0.84 | 0.87 | 0.86 | 0.91 | 0.88 |
| Case 4 | 0.91 | 0.95 | 0.93 | 0.95 | 0.90 | **0.93** |

As seen from Table 4, the RF classifier gave the best result for the Case 4 conceptual model where the $F_1$-score of Class 1 (which is our target Class) is 0.93. That means the inclusion of *question-answer data* and *product description data* significantly improved the classifier accuracy. We also calculated the individual effect of these two variables and found that the $F_1$-scores were increased from 0.86 for Case 1 as is shown in Table 3 to 0.91 and 0.88 as is shown in Table 4 for Case 2 and Case 3, respectively, for our target class. We present the confusion matrix for Case 4 in Figure 4. The confusion matrix shows how the data were classified over the two classes. We tested the Case 4 conceptual model with 12,882 reviews in which 635 high-quality reviews (target class) were wrongly classified as low-quality reviews.

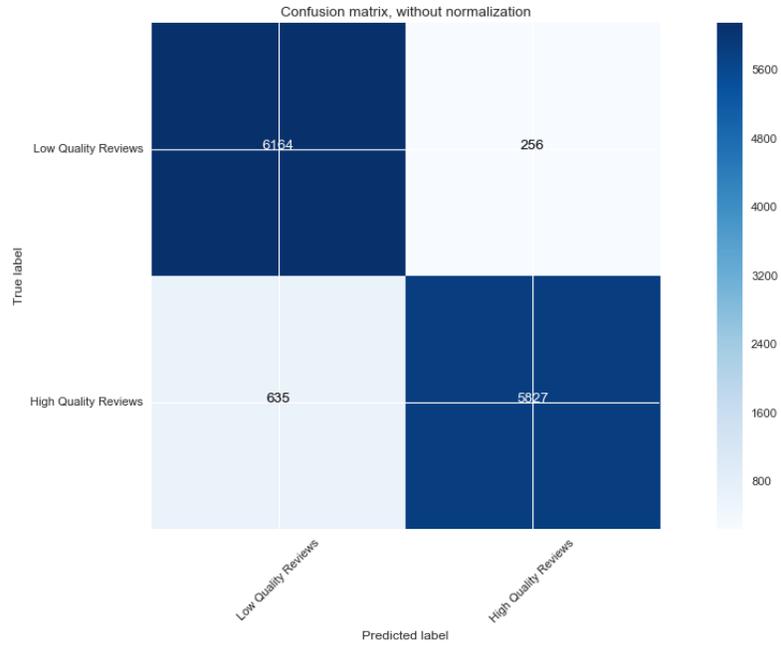

Figure 4. The confusion matrix for *Case 4 conceptual model* for Amazon test data

We also evaluated the accuracy of the RF classifier in terms of area under an ROC curve. The ROC curves for the Case 1 and Case 4 conceptual models are shown in Figure 5 and Figure 6. Figures 5 and 6 confirm that after including the *product description data* and the *customer question-answer data* as features in the feature set, the classifier accuracy increased positively from 0.87 to 0.93.

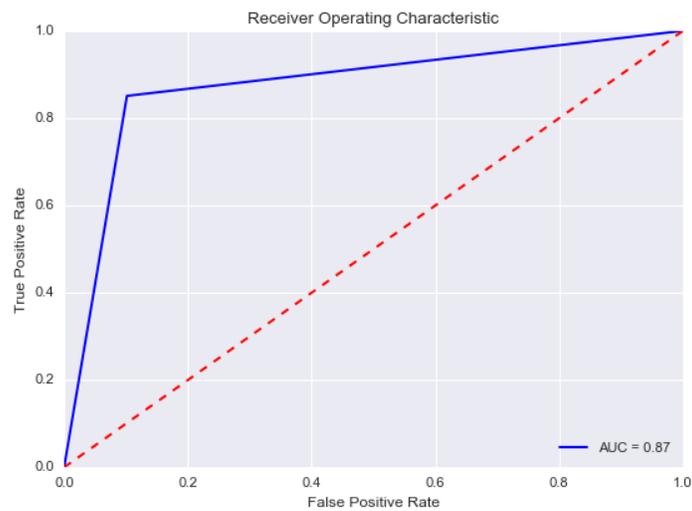

Figure 5. The ROC curve for *Baseline conceptual model* for Amazon data

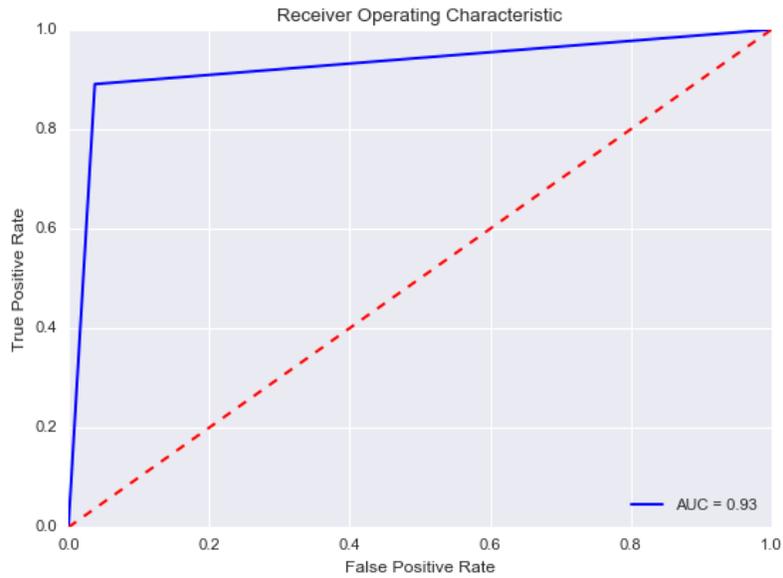

Figure 6. The ROC curve for *Case 4 conceptual model* for Amazon data

We repeated the same process for the Snapdeal dataset. Out of 10,255 Snapdeal reviews, only 494 reviews were labeled high-quality reviews and the remaining 9,761 reviews were of low quality. Hence, first we balanced the reviews over two classes using SMOTE over-sampling; then that balanced set of reviews was first divided into a 3:1 ratio for training and testing and then given as an input to the classifiers SVM, NB, and RF. The best results obtained in the testing case were for the RF classifier. The detailed results of the RF classifier for all four cases are shown in Table 5. As seen from Table 5, for the baseline model, the $F_1$-score is 0.65 for our target class, which is quite low. However, the $F_1$-score has increased from 0.65 in the baseline conceptual model to 0.85 in Case 4 where the testing feature set included the *product description data* and the *customer question-answer data*. The individual effect of these two variables was also calculated, and it was found that the $F_1$-scores increased from 0.65 for the baseline model to 0.78 and 0.74 for Case 2 and Case 3, respectively, for our target class, as shown in Table 5.

Table 5. Classification accuracy result of different cases for test data of Snapdeal

|  | Class 0 | | | Class 1 | | |
| --- | --- | --- | --- | --- | --- | --- |
|  | Precision | Recall | $F_1$-score | Precision | Recall | $F_1$-score |
| Case 1 | 0.66 | 0.75 | 0.70 | 0.70 | 0.61 | 0.65 |
| Case 2 | 0.68 | 0.70 | 0.69 | 0.82 | 0.75 | 0.78 |
| Case 3 | 0.73 | 0.71 | 0.72 | 0.76 | 0.72 | 0.74 |
| Case 4 | 0.78 | 0.78 | 0.78 | 0.84 | 0.86 | **0.85** |

*Regression:* Once the data were classified into two classes, we performed a regression. Two approaches, linear regression and gradient boosting (GB) regression (an ensemble learning technique), were used to predict the helpfulness score of the reviews. However, the performance of linear regression was quite poor compared to the GB regression. Hence, we validated our model using GB regression. The Case 4 conceptual model better classified the high-quality reviews from low-quality reviews (See Table 4 and Table 5). Thus, we used the Class 1 dataset, or the high-quality reviews obtained from the Case 4 conceptual model, to train and test the regressor. That means our regression model was built only on high-quality reviews. We used the same set of features that we used in the Case 4 conceptual model to train the classifier. The training and testing set of reviews were divided into a 3:1 ratio. The performance of the GB was measured by a mean square error. We obtained 0.267 MSE for the Amazon dataset. The MSE for the Snapdeal dataset was 0.623. Later, to rank the reviews based on their predicted helpfulness score, we simply sorted the reviews in ascending order of their votes.

Because our aim was to produce the top *k* high-quality reviews positioned at the front, we extracted the top *k* reviews from the list of high-quality reviews predicted by our system. We named this list the *"predicted top k"* list. We also extracted the top *k* actual reviews from the original or actual list of reviews that we scrapped for each product. We called this list the *"actual top k"*. To see the agreement between our *"predicted top k"* reviews with the *"actual top k"* reviews of a product, we performed the intersection between these lists as shown in Equation 5.

$$Matching = predicted\ top\ k \cap actual\ top\ k \quad (5)$$

In our case, we have taken the *k* value as 10, because people generally see the top 10 reviews before making their purchase decision (BrightLocal, 2016). Hence, Equation 5 can be further written as follows:

$$Matching = predicted\ top\ 10 \cap actual\ top\ 10 \quad (6)$$

For each product, we calculated the matching score, and we found that, for Amazon, the average matching was 5.66, whereas for Snapdeal, it was 6.12.

*4.2. Rank prediction without classifier*

To determine the impact of the classifier in our proposed model, we excluded classification phase. This means that after the feature extraction phase, we directly fed reviews (high and low quality) to the regressor. The set of reviews was given to the regressor, which predicted the helpfulness score for each review. The ranking of reviews was done by sorting the helpfulness score of each review in ascending order. We found that the MSE was increased from 0.267 to 2.545 for the Amazon dataset. The average matching between *"predicted top 10"* and *"actual top 10"* was calculated as 5.23. We repeated the same process for Snapdeal. The MSE for Snapdeal also increased from 0.623 to 3.434, and the average matching was noted as 4.32.

*4.3. Testing*

To check the performance of our system (as presented in Section 3), we additionally collected 1,000 reviews from Amazon. These data were only used for testing purposes. We passed these data to the classifier and found that only 20% (200 reviews) of the reviews were of high quality, and the rest 80% (800 reviews) were of low quality. Thus, for that 80% of the reviews, regression was not required. We simply appended those low-quality reviews at the bottom of the review list. A ratio of 20% of high-quality reviews was then fed to the regressor. Two lists *"actual top 10″* and *"predicted top 10″* were then extracted and compared. The average matching was 6.87. We also manually inspected the *"actual top ten"* reviews and the *"predicted top ten"* reviews. To achieve this, we hired 20 graduate students and gave them two lists (*predicted top 10* reviews and *actual top 10* reviews). They were asked to read the reviews of both lists and annotate which list they found more effective. In 80% of the cases, our *"predicted top 10"* reviews were annotated better than the *"actual top 10"* reviews.

## 5. Discussion

The two major highlights of our work are (i) the classification of reviews into high and low quality based on their predicted helpfulness votes and (ii) the construction of a better regression prediction model based only on high-quality reviews to rank the reviews based on their predicted votes.

Generally, when we build a regression model, we train it with both high and low-quality reviews. However, the current research first separates the high-quality reviews from the low-quality reviews and hence trains the regressor with only high-quality reviews. This means that the current research uses a hybrid model of classification and regression for finding top *k* high-quality reviews at the front. The low-quality reviews were simply added at the end of the review listing. We validated our proposed model with two different datasets from Amazon and Snapdeal. In both datasets, the reviews were highly imbalanced between the low- and high-quality classes. SMOTE over-sampling was used to balance the reviews over two classes, and those balanced reviews were then fed to the classifiers. Three different classifiers, NB, SVM and RF, were used for classifying the reviews into two classes. Our best obtained results were from the RF classifier (see Table 3). We made four conceptual models for classification, Case1, Case2, Case3 and Case4. They were different based on the sets of features that were incorporated for training and testing.

Table 6. Classification result summarization and comparison

| Approaches | Features | | | Source | Accuracy |
|---|---|---|---|---|---|
| | Textual | PDD* | CQA** | | |
| Zhang and Tran (2010) | ✓ | ✗ | ✗ | Amazon | 0.76 |
| Ghose and Ipeirotis (2011) | ✓ | ✗ | ✗ | Amazon | 0.78 – 0.87 |
| Krishnamoorthy (2015) | ✓ | ✗ | ✗ | (Blitzer, Dredze, and Pereira, 2007) and Amazon | 0.77 – 0.87 |
| Our Approach (Baseline) | ✓ | ✗ | ✗ | Amazon | 0.86 |
| Our Approach (Case 2) | ✓ | ✓ | ✗ | Amazon | 0.91 |
| Our Approach (Case 3) | ✓ | ✗ | ✓ | Amazon | 0.88 |
| Our Approach (Case 4) | ✓ | ✓ | ✓ | Amazon | **0.93** |
| Our Approach (Baseline) | ✓ | ✗ | ✗ | Snapdeal | 0.65 |
| Our Approach (Case 2) | ✓ | ✓ | ✗ | Snapdeal | 0.78 |
| Our Approach (Case 3) | ✓ | ✗ | ✓ | Snapdeal | 0.74 |
| Our Approach (Case 4) | ✓ | ✓ | ✓ | Snapdeal | **0.85** |

PDD*: Product Description data

CQA**: Customer Question-Answer data

The current study used two new features, *product description data* and *customer question-answer data*, along with the *textual features* of the reviews. The impact of the proposed features on both Amazon and Snapdeal datasets are shown in Table 6. For the Amazon dataset, in the baseline model, we used only the features from the review text and obtained an $F_1$-score of 0.86. However, the $F_1$-score was increased from 0.86 to 0.91 in Case 2 and from 0.86 to 0.88 in Case 3, in which we integrated *textual features with product description data* and *textual features with customer question-answer data*, respectively. The best result RF classifier was found for Case 4 model in which we used the features of *review text*, *product description data* and *customer question-answer data* together with a 0.93 $F_1$-score. Similarly, for the Snapdeal datasets, in the baseline model, the $F_1$-score was 0.65. It was increased in Case 2 to 0.78 and in Case 3 to 0.74. The best case was the Case 4 model in which the $F_1$-score was increased the highest from 0.65 to 0.85.

We compared our results with three earlier works in which the RF classifier was used to classify reviews into two classes (see Table 6). Zhang and Tran (2010) worked on the *textual features* derived from digital camera reviews from Amazon and obtained an $F_1$-score of 0.76. Ghose and Ipeirotis (2011) achieved an $F_1$-score of 0.78, 0.87 and 0.87 on a dataset of DVDs, audio and video, and digital cameras from Amazon. Similarly, Krishnamoorthy (2015) worked on the linguistic features derived from review text along with review metadata features to predict the helpfulness of online consumer reviews. He obtained an $F_1$-score of 0.77 and 0.87 for two datasets of reviews from Amazon and (Blitzer, Dredze, and Pereira, 2007). In line with an earlier study, the current approach performed better in classifying reviews into two classes with an $F_1$-score of 0.93 for the Amazon reviews. For the Snapdeal reviews, our results were better than all of the earlier three studies, except for that of Krishnamoorthy (2015) downloaded from (Blitzer, Dredze, and Pereira, 2007). However, most importantly, in the aforementioned works, they set forth their classification results as final results. However, in our case, the classification results served only as a partial result, as our main objective was to build a better regression model that would be trained only on high-quality reviews. To determine how classification before regression effects overall accuracy, we performed experiments in two phases: i) training the regressor only on high-quality reviews (Section 4.1), and ii) training the regressor with both high and low-quality reviews (Section 4.2). We found that MSE was 0.267 when trained only on high-quality reviews; however, it increased to 2.545 when trained on both low- and high-quality reviews from Amazon. Similarly, for Snapdeal, the MSE was 0.623 when trained only on high-quality reviews; however, it increased to 3.434 when trained with both classes. We also evaluated the impact of the *product description data* and the *customer question answer data* on positive and negative high-quality reviews separately. We found that negative reviews (with star ratings of 1 or 2 or 3) were not affected too much by the *product description data*, whereas positive reviews (with star ratings of 4 or 5), which included texts of *product description data* in the review texts, were highly voted. Similarly, for negative reviews, the inclusion of *customer question-answer data* did not have much of an effect in terms of votes, whereas positive reviews that included the topics of the *customer question-answer data* were highly affected.

After the regression, we sorted and ranked the reviews based on their predicted helpfulness values. To determine the agreement between the proposed system and the prevailing system on Amazon and Snapdeal, two lists *"predicted top k"* reviews and *"actual top k"* reviews were extracted. We took the k value as 10; hence, the *predicted top 10* and *actual top 10* were extracted. In the existing system, all the top reviews are based only on real votes obtained by those reviews; however, in the current system, we ranked them based on their quality. Thus, when we matched the existing system with our proposed one, we found that for both cases, Amazon and Snapdeal, 5-to-6 reviews are the same. However, the remaining 4-to-5 reviews in our system are new reviews, which had not obtained sufficient votes. These reviews are highly salient; however, because they were posted late on the product page, they did not receive the exposure. We validated the quality of our predicted reviews by human evaluators. They confirmed that our top 10 reviews are better than the prevailing reviews of Amazon and Snapdeal. The previous studies Zhang and Tran (2010) and Singh et al. (2017a) proposed models to automatically predict the helpfulness of online consumer reviews using only review text, and they claimed to rank those reviews. However, they did not reveal a mechanism for placing the reviews in the proper place nor did they obtain agreement between their system and the existing systems. In line with those earlier works, the presented technique can be considered as an extension.

*5.1 Implications for theory*

This paper expands the flourishing literature on the usefulness of online reviews in two ways. First, it supplements the prior research (Singh et al., 2017a) by revealing the mechanism to rank the reviews on the product page. Here, we considered only high-quality reviews while predicting top ten reviews. One of the major contributions of this research is data balancing before classification to improve the accuracy of the classifier. The current research is the first of its type to separate high-quality reviews from low-quality reviews using a random forest classifier. The filtered reviews are then placed in a suitable position in the review list based on predicted helpfulness votes. The removal of low-quality reviews improves the time complexity as there will be no change of list for the low-quality reviews. For simplicity and clarity, we have shown the experiments for only top ten predicted reviews and compared them with actual top ten reviews. However, the presented technique considers them all to be high-quality reviews, finds their initial helpfulness votes and ranks them accordingly. Second, this paper identifies two new qualitative review features, specifically, *product description* and *customer question-answer*, which to the best of our knowledge have not previously been explored. Both *product description* and *customer question-answer* showed a positive association with review helpfulness. Moreover, for negative reviews, the effect of *product description data* on helpfulness was found to be neutral. A deeper analysis of such negative reviews is required to reveal the exact reason for their neutrality. With the current form of research, we used a similarity measure between *product description* and review text as well as between *customer question-answer* and review text. In the future, researchers can employ other measures to enhance the importance of said features offer a comprehensive understanding of review helpfulness.

*5.2 Implications for practice:*

One of the major implications of this research is that reviewers are guided to add the topics of *product description data* and *question-answer data* to their studies. Website developers who use this research to redesign review pages may wish to include these topics in their review pages. The ranking becomes dynamic as even new reviews may be inserted at top, which will encourage reviewers to write better reviews. The inclusion of this algorithm does not change the website layout as this classification and regression model will run in the background. Better reviews will be available to be positioned on top, which will engage more customers and encourage them to write high-quality reviews. The proposed system can be implemented on top of existing systems for any e-commerce site or review rating site. Top quality reviews can be placed in a separate linked list to facilitate easier manipulation of review insertion and rearrangement. Thus, for software developers, it will be quite easy to implement this system.

## 6. Conclusion, Limitations and Future Research

The paper proposes an automated system that assigns an initial helpfulness vote to a review as soon it has been posted on the product page and uses this vote to place it at the proper place in the review listing. This study first separates the high-quality reviews from low-quality ones and hence predicts the top 10 reviews from high-quality reviews only. Low-quality reviews are simply appended at the end of the review queue. The main contribution of this research is to create an opportunity for all reviews to be seen at the top of the list of reviews based on their predicted vote, regardless of their posting date. Apart from this, the paper seeks to answer three research questions: (1) will excluding low-quality reviews improve regressor performance? (2) Does the inclusion of *product description data* in a review affect the helpfulness of online consumer reviews? If so how? (3) Does the inclusion of topics on *question-answer data* affect the helpfulness of online consumer reviews? If so how? Drawing on data from Snapdeal and Amazon, it was found that *product description data* is an important parameter for the helpfulness of positive reviews while having a neutral effect on review helpfulness for critical or negative reviews. Similarly, *question-answer data* showed a positive association with review helpfulness.

However, as with any other research, this research is also not without limitations. First, the dataset was constrained by the choice of Snapdeal and Amazon as the data sources. Hence, the outcomes of this research should not be considered to be generalized and should be carefully implemented in different contexts. Second, the data collection for this study was completed for only one window period. An awareness is therefore required when aligning the current findings to data collected at different time periods and from different sources such as flipkart.com or yelp.com. Third, the findings of this study are significant only to the chosen five product categories. Further research is needed to validate whether they are generalizable to other product categories such as fashion products or sports products. Fourth, the technique used for ranking the online reviews is restricted with regard to implementing the techniques employed, which can be further improved by future research in terms of using other regression and classification techniques. Further, a deep learning type structure can also be employed, which will remove the effect of biased hand-crafted features on the helpfulness prediction of reviews and can itself be trained with raw data. The system performance can be further improved by using certain extra features such as product sales rank, or review comments. Review listings may also be improved by giving more weight to recent high-quality reviews over older reviews.

**Acknowledgments.**

The author would like to acknowledge the Ministry of Electronics and Information Technology (MeitY), Government of India for the financial support during research work through "Visvesvaraya PhD Scheme for Electronics and IT".

**References**

Allahbakhsh, M., Ignjatovic, A., Motahari-Nezhad, H.R. and Benatallah, B., 2015. Robust evaluation of products and reviewers in social rating systems. World Wide Web, 18(1), pp.73-109.

Baek, H., Lee, S., Oh, S. and Ahn, J., 2015. Normative social influence and online review helpfulness: Polynomial modeling and response surface analysis. Journal of Electronic Commerce Research, 16(4), p.290.

Blitzer, J., Dredze, M. and Pereira, F., 2007. Biographies, bollywood, boom-boxes and blenders: Domain adaptation for sentiment classification. In Proceedings of the 45th annual meeting of the association of computational linguistics (pp. 440-447).

BrightLocal, 2016. Local consumer review survey accessed from www.brightlocal.com/learn/local-consumer-review-survey/ Accessed on 22nd December 2016.

Chawla, N.V., Bowyer, K.W., Hall, L.O. and Kegelmeyer, W.P., 2002. SMOTE: synthetic minority over-sampling technique. Journal of artificial intelligence research, 16, pp.321-357.

Chen, H.N. and Huang, C.Y., 2013. An investigation into online reviewers' behavior. European Journal of Marketing, 47(10), pp.1758-1773.

Chua, A.Y. and Banerjee, S., 2015. Understanding review helpfulness as a function of reviewer reputation, review rating, and review depth. Journal of the Association for Information Science and Technology, 66(2), pp.354-362.

Chua, A.Y. and Banerjee, S., 2016. Helpfulness of user-generated reviews as a function of review sentiment, product type and information quality. Computers in Human Behavior, 54, pp.547-554.

Davis, J. and Goadrich, M., 2006, June. The relationship between Precision-Recall and ROC curves. In Proceedings of the 23rd international conference on Machine learning (pp. 233-240). ACM.

Ghose, A. and Ipeirotis, P.G., 2011. Estimating the helpfulness and economic impact of product reviews: Mining text and reviewer characteristics. IEEE Transactions on Knowledge and Data Engineering, 23(10), pp.1498-1512.

Ghose, A. and Ipeirotis, P.G., 2007, August. Designing novel review ranking systems: predicting the usefulness and impact of reviews. In Proceedings of the ninth international conference on Electronic commerce (pp. 303-310). ACM.

Heinonen, K., 2011. Consumer activity in social media: Managerial approaches to consumers' social media behavior. Journal of Consumer Behaviour, 10(6), pp.356-364.

Huang, A.H., Chen, K., Yen, D.C. and Tran, T.P., 2015. A study of factors that contribute to online review helpfulness. Computers in Human Behavior, 48, pp.17-27.

Karmaker Santu, S.K., Sondhi, P. and Zhai, C., 2016, October. Generative feature language models for mining implicit features from customer reviews. In Proceedings of the 25th ACM International on Conference on Information and Knowledge Management (pp. 929-938). ACM.

Kim, S.M., Pantel, P., Chklovski, T. and Pennacchiotti, M., 2006, July. Automatically assessing review helpfulness. In Proceedings of the 2006 Conference on empirical methods in natural language processing (pp. 423-430). Association for Computational Linguistics.

Korfiatis, N., GarcíA-Bariocanal, E. and SáNchez-Alonso, S., 2012. Evaluating content quality and helpfulness of online product reviews: The interplay of review helpfulness vs. review content. Electronic Commerce Research and Applications, 11(3), pp.205-217.

Krishnamoorthy, S., 2015. Linguistic features for review helpfulness prediction. Expert Systems with Applications, 42(7), pp.3751-3759.

Lee, E.J. and Shin, S.Y., 2014. When do consumers buy online product reviews? Effects of review quality, product type, and reviewer's photo. Computers in Human Behavior, 31, pp.356-366.

Lee, S. and Choeh, J.Y., 2014. Predicting the helpfulness of online reviews using multilayer perceptron neural networks. Expert Systems with Applications, 41(6), pp.3041-3046.

Li, M., Huang, L., Tan, C.H. and Wei, K.K., 2013. Helpfulness of online product reviews as seen by consumers: Source and content features. International Journal of Electronic Commerce, 17(4), pp.101-136.

Liu, J., Cao, Y., Lin, C.Y., Huang, Y. and Zhou, M., 2007. Low-quality product review detection in opinion summarization. In Proceedings of the 2007 Joint Conference on Empirical Methods in Natural Language Processing and Computational Natural Language Learning (EMNLP-CoNLL).

Liu, Y., Huang, X., An, A. and Yu, X., 2008, December. Modeling and predicting the helpfulness of online reviews. In Data mining, 2008. ICDM'08. Eighth IEEE international conference on (pp. 443-452). IEEE.

Mackiewicz, J. and Yeats, D., 2014. Product review users' perceptions of review quality: The role of credibility, informativeness, and readability. IEEE Transactions on Professional Communication, 57(4), pp.309-324.

Malik, M.S.I. and Hussain, A., 2017. Helpfulness of product reviews as a function of discrete positive and negative emotions. Computers in Human Behavior, 73, pp.290-302.

Merton, R.K., 1968. The Matthew effect in science: The reward and communication systems of science are considered. Science, 159(3810), pp.56-63.

Mudambi, S.M. and Schuff, D., 2010. What makes a helpful online review? A study of customer reviews on Amazon. com. MIS quarterly, pp.185-200.

Ngo-Ye, T.L. and Sinha, A.P., 2014. The influence of reviewer engagement characteristics on online review helpfulness: A text regression model. Decision Support Systems, 61, pp.47-58.

Qazi, A., Syed, K.B.S., Raj, R.G., Cambria, E., Tahir, M. and Alghazzawi, D., 2016. A concept-level approach to the analysis of online review helpfulness. Computers in Human Behavior, 58, pp.75-81.

Racherla, P. and Friske, W., 2012. Perceived 'usefulness' of online consumer reviews: An exploratory investigation across three services categories. Electronic Commerce Research and Applications, 11(6), pp.548-559.

Roy, P.K., Ahmad, Z., Singh, J.P., Alryalat, M.A.A., Rana, N.P. and Dwivedi, Y.K., 2018. Finding and Ranking High-Quality Answers in Community Question Answering Sites. Global Journal of Flexible Systems Management, 19(1), pp.53-68.

Saini, S., Saumya, S. and Singh, J.P., 2017, June. Sequential Purchase Recommendation System for E-Commerce Sites. In IFIP International Conference on Computer Information Systems and Industrial Management (pp. 366-375). Springer.

Saumya, S., Singh, J.P. and Kumar, P., 2016, September. Predicting Stock Movements using Social Network. In Conference on e-Business, e-Services and e-Society (pp. 567-572). Springer.

Siering, M. and Muntermann, J., 2013. What Drives the Helpfulness of Online Product Reviews? From Stars to Facts and Emotions. In Wirtschaftsinformatik (Vol. 7).

Singh, J.P., Irani, S., Rana, N.P., Dwivedi, Y.K., Saumya, S. and Roy, P.K., 2017a. Predicting the "helpfulness" of online consumer reviews. Journal of Business Research, 70, pp.346-355.

Singh, J.P., Dwivedi, Y.K., Rana, N.P., Kumar, A. and Kapoor, K.K., 2017b. Event classification and location prediction from tweets during disasters. Annals of Operations Research, pp.1-21.

Tsang, A.S. and Prendergast, G., 2009. Is a "star" worth a thousand words? The interplay between product-review texts and rating valences. European Journal of Marketing, 43(11/12), pp.1269-1280.

Weathers, D., Swain, S.D. and Grover, V., 2015. Can online product reviews be more helpful? Examining characteristics of information content by product type. Decision Support Systems, 79, pp.12-23.

Zhang, R. and Tran, T., 2010. Helpful or unhelpful: a linear approach for ranking product reviews. Journal of Electronic Commerce Research, 11(3), p.220.

Zheng, X., Zhu, S. and Lin, Z., 2013. Capturing the essence of word-of-mouth for social commerce: Assessing the quality of online e-commerce reviews by a semi-supervised approach. Decision Support Systems, 56, pp.211-222.

Zhu, L., Yin, G. and He, W., 2014. Is this opinion leader's review useful? Peripheral cues for online review helpfulness. Journal of Electronic Commerce Research, 15(4), pp.267-280.